\documentclass[oldversion]{aa}
\usepackage{graphicx}
\usepackage{txfonts}
\usepackage{natbib}
\usepackage{url}
\usepackage{multirow}
\usepackage{hyperref}

\begin{document}
   \title{Non-LTE calculations for neutral Na in late-type stars using
     improved atomic data}

   \author{K. Lind\inst{1,2}\and
           M. Asplund\inst{2}\and
           P.\,S.\ Barklem\inst{3}\and
           A.\,K.\ Belyaev\inst{4}} 

   \institute{European Southern Observatory (ESO), Karl-Schwarzschild-Strasse 2,
              85748 Garching bei M\"unchen, Germany
             \and
              Max-Planck-Institut f\"ur Astrophysik, Karl-Schwarzschild-Strasse 1,
              85741 Garching bei M\"unchen, Germany\\
               \email{klind@mpa-garching.mpg.de}\and
              Department of Physics \& Astronomy, Uppsala University, Box 515, 75120 Uppsala, Sweden\and
              Department of Theoretical Physics, Herzen University, St. Petersburg 191186, Russia
             }
   
   \date{Received 08/11/2010, Accepted 19/01/2011}

\abstract{Neutral sodium is a minority species in the atmospheres of late-type
  stars, and line formation in local thermodynamic equilibrium (LTE) is often
  a poor assumption, in particular for strong lines. We present an
  extensive grid of non-LTE calculations for several Na\,I lines in
  cool stellar atmospheres, including metal-rich and metal-poor dwarfs
  and giants. For the first time, we constructed a Na model
  atom that incorporates accurate quantum mechanical calculations
  for collisional excitation and ionisation by electrons as well as
  collisional excitation and charge exchange reactions with neutral
  hydrogen. Similar to Li\,I, the new rates for hydrogen
    impact excitation do
  not affect the statistical equilibrium calculations,
  while charge exchange reactions have a small
  but non-negligible influence.
  The presented LTE and non-LTE
  curves-of-growth can be interpolated to obtain non-LTE abundances
  and abundance corrections for arbitrary stellar parameter
  combinations and line strengths. The
  typical corrections for weak lines are $-0.1...-0.2$\,dex,
  whereas saturated lines may overestimate the abundance in LTE by
  more than 0.5\,dex. The non-LTE Na abundances appear very robust with
  respect to uncertainties in the input collisional data.}

   \keywords{Stars: abundances -- Stars: late-type -- Line: formation }

   \titlerunning{Non-LTE calculations for neutral Na}
   \maketitle

\section{Introduction}
\label{sec:intro}
Sodium has established itself as an important tracer of Galactic chemical
evolution, and numerous investigations of the Na abundances of
late-type stars, residing in different regions of the Galaxy, have
been conducted
\citep[e.g.][]{Takeda03,Gehren06,Andrievsky07}. Na is mainly
synthesised during hydrostatic carbon-burning in massive stars,
through the reaction $^{12}$C($^{12}$C,p)$^{23}$Na. As pointed out by
\citet{Woosley95}, the production is dependent on the available
neutron excess through secondary reactions, which implies
metal-dependent yields. In addition, there is a production channel via
proton-capture reactions, $^{22}$Ne(p,$\gamma$)$^{23}$Na
\citep{Denisenkov90}. The latter, the so-called NeNa-cycle, occurs when
temperatures are high enough for H-burning through the CNO-cycle,
e.g. in the cores or H-burning shells of intermediate mass and
massive stars. 

Abundance studies of late-type stars in the thin disk show a steady
increase from solar to positive [Na/Fe] ratios at super-solar
metallicities, while thin and thick disk stars instead form a
decreasing trend slightly below solar metallicities
\citep[e.g.][]{Edvardsson93,Reddy03,Bensby03,Shi04}. Relying
exclusively on weak lines for the analysis, LTE has been proved a
reasonable approximation in this metallicity regime. For metal-poor
stars the situation is different, especially in cases where the strong
Na\,I D resonance lines (588.9/589.5\,nm) are the only available
abundance indicators. As shown by \citet{Takeda03} and
\citet{Gehren06}, [Na/Fe] ratios are slightly subsolar
($-0.1...-0.5$) in metal-poor stars in the thick disk and the halo in
the metallicity range [Fe/H]$=-3.0...-1.0$. This deficiency is only
recovered through non-LTE analysis, because LTE investigations tend to
overestimate the abundances, sometimes by as much as $\sim0.5$\,dex
(see Sect.\,\ref{sec:deps}). Moreover, almost solar values are obtained from non-LTE
analysis of extremely metal-poor stars below $\rm[Fe/H]<-3.0$, where
LTE analysis, at least of giants, instead yields positive ratios,
[Na/Fe]$\approx0.3$ \citep{Cayrel04,Andrievsky07}. Interestingly,
\citet{Nissen10} found evidence for systematic Na abundance
differences of the order of 0.2\,dex between $\alpha$-poor and $\alpha$-rich
halo stars, with important implications for the presumably separate
origin of these two Galactic components. To place all Galactic stellar
populations on an absolute Na abundance scale to the same and better
precision, non-LTE is clearly required.  

In globular clusters, Na is of particular interest, because the large
overabundances of this element, compared with those of field stars of similar
metallicities, imply a chemical evolution scenario that is specific to
these dense stellar systems \citep[e.g.][]{Gratton04}. By detailed
mapping of the Na abundance and its correlating behaviour with
similar-mass and lighter elements we may distinguish between stars
formed in different formation episodes in globular clusters and
eventually identify the elusive self-enrichment process that so
efficiently polluted the star-forming gas with the nucleosynthesis
products of hot H-burning through the CNO-cycle and the related NeNa-
and MgAl-chains (i.e. enhancement of N, Na, and Al, and depletion of
C, O, and Mg, see e.g. \citealt{Lind10a} and \citealt{Carretta09b}).

In this study we present 1D, non-LTE calculations for several neutral
Na lines for a large stellar grid. The given non-LTE abundances
can be interpolated to arbitrary stellar-parameter
combinations and will consequently be useful for Na abundance analyses with
a variety of applications. In a forthcoming paper we will apply the
non-LTE modelling procedure described here to individual stars, and
extend it to using 3D hydrodynamical model atmospheres. 

Typically, the largest uncertainties affecting the non-LTE
calculations in stellar atmospheres are collisional cross-sections,
especially for collisions with hydrogen atoms
\citep[e.g.][]{Asplund05}. As described in Sect.\,\ref{sec:ecoll} and
\ref{sec:hcoll}, recent
quantum mechanical calculations of cross-sections for collisions with
both electrons and hydrogen have significantly improved the situation
for sodium, and a reliable atom can now be constructed. In this study,
we put extra emphasis on assessing to which extent the remaining
uncertainties in collisional rates influence the statistical
equilibrium of Na\,I.

\section{Non-LTE modelling procedure}
\label{sec:modelling}
We used the code MULTI, version 2.3 \citep{Carlsson86,Carlsson92} to
simultaneously solve the statistical equilibrium and radiative
transfer problems in a plane-parallel stellar atmosphere. Na was
considered a trace element, i.e. we neglected feedback effects from changes
in its level populations on the atmospheric structure. The LTE
assumption was used for all other species in the computation of
background continuum and line opacity. A simultaneously computed
line-blanketed radiation field was therefore used in the calculation of
photoionisation rates. 

A grid of 764 1D, LTE, opacity-sampling \textsc{MARCS} model
atmospheres \citep{Gustafsson08} was used 
in the analysis. The models span $T_{\rm eff}=4000...8000$\,K,
$\log{g}=1.0...5.0$, and $\rm[Fe/H]=-5.0...+0.5$. Sodium abundances
vary from $\rm[Na/Fe]=-2.0...+2.0$, in steps of 0.25\,dex. The highest
effective temperature is 5500\,K for models with $\log{g}=1.0$,
6500\,K for $\log{g}=2.0$, and 8000\,K for
$\log{g}\ge3.0$. For models with $\log{g}\ge3.0$, we adopted 
microturbulence parameters $\xi_{\rm t}=1.0$ and $2.0\rm\,km\,s^{-1}$
and for models with $\log{g}\le3.0$, we adopted 
$\xi_{\rm t}=1.0,2.0,$ and $5.0\rm\,km\,s^{-1}$. All models have a
standard composition,
i.e. with scaled solar abundances according to \citet{Grevesse07},
plus 0.4\,dex enhancement of alpha-elements in all metal-poor models
($\rm[Fe/H]\le-1.0$).

We define a non-LTE correction for each abundance point as the
difference between the LTE sodium abundance and the non-LTE abundance
that corresponds to the same equivalent width. Corrections are given
for equivalent widths in the range $0.01-100$\,pm. The equivalent
width was obtained by numerical integration over the line profile,
considering a spectral region that extends $\pm0.3-3\rm\,nm$ from the
line centre, depending on the typical line strength. 


\subsection{Structure of the model atom}
\label{sec:atom}
\begin{figure}
        \centering
                \includegraphics[angle=90,width=9.0cm]{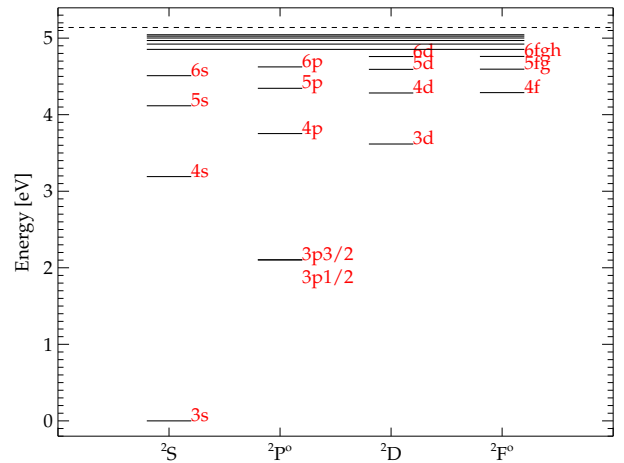}
          \caption{Schematic term diagram of the 23-level Na\,I model
            atom. The dashed line marks the Na\,II continuum. The
            highest level considered in Na\,I is $n=12$.}
        \label{fig:term}
\end{figure}

The model atom we constructed for Na consists of 22 energy levels
of Na\,I, plus the Na\,II continuum (see Fig.\,\ref{fig:term}). The
energy levels are coupled by radiative transitions and by
collisional transitions with electrons and neutral
hydrogen. Experimentally measured energies were taken from the
compilation by \citet{Sansonetti08} (for highly excited states
isoelectronic fitting was used). Because we are not concerned with
modelling the detailed structure of highly excited levels, we
collapsed all sublevels for $n=7-12$ into super-levels. The
fine-structure components of the 3p level were accounted for in the
statistical equilibrium calculations by treating $\rm 3p_{1/2}$ and
$\rm 3p_{3/2}$ as separate, collisionally coupled
levels\footnote{Accounting for the fine-structure components of the 3p
  level simplifies the numerical procedure, while also predicting
  correct line formation depths for the individual lines
  \citep[e.g.][]{Mashonkina00}.}. In addition, the fine-structure
components of the 3p--3d and 3p--4d transitions, as well as the
hyper-fine structure components of the 3s--$\rm 3p_{1/2}$ and 3s--$\rm
3p_{3/2}$ transitions, were accounted for by computing the line profile
as a linear combination of the subcomponents (see Table
\ref{tab:lines}). 

The collisional transition probabilities were computed for the 3p level,
not for its fine-structure components. To calculate these we assume
that the ionisation and excitation rates of the sublevels were equal
to that of the collapsed level. We further assume that the
de-excitation rates to 3s scale with the statistical weights, i.e. the
de-excitation rate from $\rm 3p_{3/2}$ is twice as large as from $\rm
3p_{1/2}$, and the sum of the rates is equal to that of the collapsed
level. The collisional cross-section for electron impact excitation
between the two fine-structure levels was estimated with the 
impact parameter method of \citet{Seaton62a}. We note that the detailed
rates are not important for the non-LTE problem.

\subsection{Radiative transitions}
\label{sec:rad}
In total, 166 allowed bound-bound radiative transitions were included,
adopting where possible oscillator strengths from the \textit{ab
  initio} calculations of C. Froese
Fischer \footnote{Multi-configuration Hartree-Fock computations
  (MCHF). \url{http://www.vuse.vanderbilt.edu/~cff/mchf_collection/}}. For
most remaining transitions, we used the calculated transition
probabilities by K.T. Taylor, as part of the Opacity
Project \citep[TOPbase][]{Cunto92} 
\footnote{\url{http://legacy.gsfc.nasa.gov/topbase}}. The two sets of
$f$-values typically agree within 3\% for the strongest, most
important transitions. For the Na\,D lines, accurate experimental data
exist, and we adopted the values listed in the NIST data
base\footnote{\url{http://physics.nist.gov/PhysRefData/ASD/index.html}}
(see Table \ref{tab:lines}).  

Photoionisation cross-sections for levels with $l\le4$ were drawn from
the TOP-base (computations by K.\,T.\,Taylor). For the highly excited
collapsed levels we adopted hydrogenic cross-sections.

\begin{table*}
      \caption{Line data for the 11 lines considered in the spectrum synthesis.}
         \label{tab:lines}
         \centering
         \begin{tabular}{lrrlllllllll}
                \hline\hline
nl--n'l' & $\lambda$ &$^{(a)}\Delta E$    &   $J$  & $J'$&  $F$  &
$F'$  &  $f$ & $^{(b)}\Gamma$ & $^{(c)}\sigma$ & $^{(d)}\alpha$  &
$^{(e)}\log{C_4}$\\
         & [nm]   &$\rm[cm^{-1}]$    &              &            &               &              &      &        $\rm[rad\,s^{-1}]$            &    [a.u.\,]         &    &             \\
\hline
\multirow{4}{*}{3s--3p} & \multirow{4}{*}{589.5} & -0.033  & 1/2  &1/2  &1  & 1   & $2.67\times 10^{-2}$  & \multirow{4}{*}{$6.14\times 10^{7}$} & \multirow{4}{*}{407} & \multirow{4}{*}{0.237} & \multirow{4}{*}{$-15.11$}\\
&& -0.039  & 1/2  &1/2  &1  & 2   & $1.33\times 10^{-1}$   \\
&&  0.026  & 1/2  &1/2  &2  & 1   & $8.00\times 10^{-2}$   \\
&&  0.020  & 1/2  &1/2  &2  & 2   & $8.00\times 10^{-2}$   \\
\hline
\multirow{2}{*}{3s--3p} & \multirow{2}{*}{588.9} & -0.036  & 1/2  &3/2  &1  & 0,1,2& $3.21\times 10^{-1}$  & \multirow{2}{*}{$6.16\times 10^{7}$} & \multirow{2}{*}{407} & \multirow{2}{*}{0.237} & \multirow{2}{*}{$-15.11$}\\
&&  0.022  & 1/2  &3/2  &2  & 1,2,3 & $3.21\times 10^{-1}$ \\
             \hline
3p--3d & 818.3 &   & 1/2  &3/2  &   &   & $8.63\times 10^{-1}$  & $1.13\times 10^8$  &  804 & 0.270 & $-13.76$\\
             \hline
\multirow{2}{*}{3p--3d} & \multirow{2}{*}{819.4} & -0.020  & 3/2  &3/2  &   &   & $8.63\times 10^{-2}$  & \multirow{2}{*}{$1.13\times 10^{8}$} & \multirow{2}{*}{804} & \multirow{2}{*}{0.270} & \multirow{2}{*}{$-13.76$}\\
 &  & 0.030  & 3/2  &5/2  &   &   & $7.77\times 10^{-1}$  \\
             \hline
3p--4d & 568.2 &   & 1/2  &3/2  &   &   & $9.82\times 10^{-2}$  & $8.05\times 10^7$  &  1955 & 0.327 & $-12.18$\\
             \hline
\multirow{2}{*}{3p--4d} & \multirow{2}{*}{568.8} & -0.014  & 3/2  &3/2  &   &   & $9.82\times 10^{-3}$  & \multirow{2}{*}{$8.07\times 10^{7}$} & \multirow{2}{*}{1955} & \multirow{2}{*}{0.327} & \multirow{2}{*}{$-12.18$}\\
 &  & 0.021  & 3/2  &5/2  &   &   & $8.84\times 10^{-2}$  \\
             \hline
3p--5s & 615.4 &   & 1/2  &1/2  &   &   & $1.42\times 10^{-2}$  & $7.43\times 10^7$  &  - & - & $-13.20$\\
             \hline
3p--5s & 616.1 &   & 3/2  &1/2  &   &   & $1.42\times 10^{-2}$  & $7.45\times 10^7$  &  - & - & $-13.20$\\
             \hline
3p--6s & 514.8 &   & 1/2  &1/2  &   &   & $4.52\times 10^{-3}$  & $6.81\times 10^7$  &  - & - & $-12.55$\\
             \hline
3p--7s & 475.1 &   & 3/2  &1/2  &   &   & $2.09\times 10^{-3}$  & $6.31\times 10^7$  &  - & - & $-10.24$ \\
             \hline
4s--5p & 1074.6&   & 1/2  &3/2  &   &   & $2.54\times 10^{-2}$  & $2.91\times 10^7$  &  - & - & $-12.64$\\
             \hline
\\
\multicolumn{12}{p{15cm}}{$^{(a)}$ $\Delta E$ is the difference in energy between the given subcomponent and the average energy gap bridged by the transition. The energy separations of the hyperfine components of 3s--3p are taken from \citet{Sydoryk08}}\\
\multicolumn{12}{p{15cm}}{$^{(b)}$ $\Gamma$ is the natural broadening
  width (FWHM).}\\
\multicolumn{12}{p{15cm}}{$^{(c)}$ $\sigma$ is the broadening cross-section for collisions with neutral hydrogen at relative velocity $v=10^4\rm m\,s^{-1}$.}\\
\multicolumn{12}{p{15cm}}{$^{(d)}$ $\alpha$ is the velocity dependence
  of $\sigma$.}\\
\multicolumn{12}{p{15cm}}{$^{(e)}$ $C_4$ in [$\rm cm^4 s^{-1}$] }
\end{tabular}
\end{table*}

\subsection{Collisions with electrons}
\label{sec:ecoll}
Cross-sections for collisional excitation by electrons can be
estimated using general recipes based on the Born
approximation, which is known to overestimate the
cross-sections at low impact energies \citep[e.g][]{Park71}. Those
near-threshold energies are most relevant for stellar atmospheres
that host electrons with typical kinetic energies of the order of
1\,eV. \citet{Seaton62a} tried to rectify the Born cross-sections by
modifying the transition probability for low-impact parameters, thus
accounting for strong coupling between states, which was previously
neglected (the so-called impact parameter approximation). Another
approach is to empirically adjust the Born rates to reach better
agreement with experimental data, which was advocated by
\citet{VanRegemorter62} and \citet{Park71}. Without any 
alternatives, non-LTE applications have long had to rely almost
exclusively on these simple semi-empirical formulae. 

\begin{figure}
        \centering
                \includegraphics[angle=90,width=8.5cm]{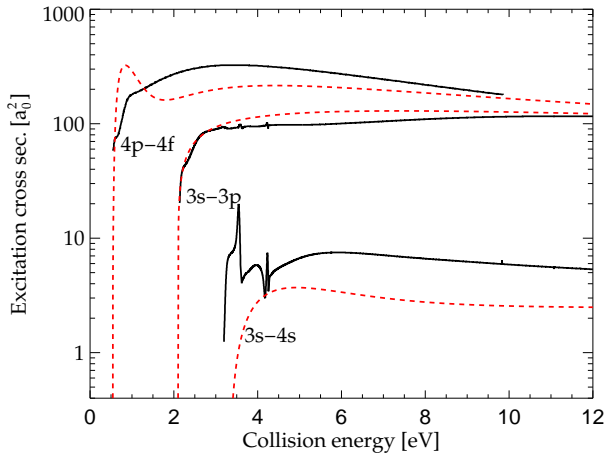}
          \caption{Three sample excitation cross-sections for
            collisions between Na\,I and electrons. The black solid
            lines represent R-matrix calculations (Feautrier et al. 2010,
            Gao et al. 2010) and the red dashed lines represent the
            analytical fitting functions derived by
            \citet{Igenbergs08}. The transitions are marked with
            labels next to the energy thresholds.}
        \label{fig:ecoll}
\end{figure}

Nowadays, much more rigorous quantum mechanical calculations can be
performed for simple atoms like Na. \citet{Igenbergs08} present new
convergent close-coupling (CCC) calculations for excitation and
ionisation of neutral Na by electron collisions, which agree well
with experimental data when available \citep[e.g. with the
  measurements by][]{Phelps81}. Additionally, R-matrix
calculations that are able to recover the detailed
resonant structures of the cross-sections at threshold impact energies
($<5$\,eV) have been performed by \citet{Gao10}. Here, we adopted
cross-sections calculated with the
R-matrix method in a manner similar to that described in Gao et
al.\,, including seven real spectroscopic states of Na\,I
(3s, 3p, 4s, 3d, 4p, 4d, and 4f) and four polarised pseudo-orbitals
to take into account the polarisation effects between the scattered
electron and the target electrons (see Feautrier et al.\,
2010, in preparation). The calculations were performed for  
impact energies in the range 0--14\,eV. Figure \ref{fig:ecoll} compares the
cross-sections obtained for some sample transitions with the R-matrix
method and the analytical fitting functions by
\citet{Igenbergs08}. The two methods agree very well; the typical
difference between calculated rate coefficients in the temperature
regime $2000-20000$\,K is only a factor of two, although individual
rates could differ by up to a factor of ten. 

For more highly excited states of Na\,I, quantum mechanical data are
lacking and we are still dependent on general 
formulae. However, the empirically adjusted rates by \citet{Park71}
perform surprisingly well compared with the newer data, with half
of the transitions at 6000\,K agreeing within a factor of three and
90\% of the transitions agreeing to within a factor of ten. We adopted
these for transitions involving levels above 5s.

Cross-sections for ionisation by electron impact have also been 
calculated with the CCC method up to 5s by \citet{Igenbergs08}, and
compare well with existing experimental data for the ground and first
excited states. For ionisation from more highly excited levels, we relied
on the general recipe given in \citet{Allen00}. Also, to bridge the
small energy gap between the ionisation edge and the minimum energy
for which the analytical fits to the ionisation cross-sections are
valid (see Table 2 in Igenbergs et al.\,), we assumed that the
cross-sections for all levels were given by the general recipe
\citep{Allen00} at the edge.

In Sect.\,\ref{sec:influence} we evaluate the influence from
uncertainties in
collisional rates on the statistical equilibrium of Na\,I by
multiplying and dividing all electron collisional rates by a factor of
two, which given the present data may be indicative of the true
uncertainty of the data for the most important transitions.
   
\subsection{Collisions with hydrogen} \ \\
\label{sec:hcoll}
Neutral hydrogen atoms are the most abundant atomic species in
late-type stellar atmospheres. Still, collisional cross-sections
between hydrogen and other atoms are notorious sources of uncertainty
for non-LTE applications, and some empirical calibration of the
classical recipe by \citet{Drawin68} is usually the only option, using
a general rate scaling factor $\rm S_H$. Again, the situation for Na
today is improved in this respect.

\begin{figure}
        \centering
                \includegraphics[angle=90,width=9.0cm,viewport=0cm 10cm 19cm 26cm]{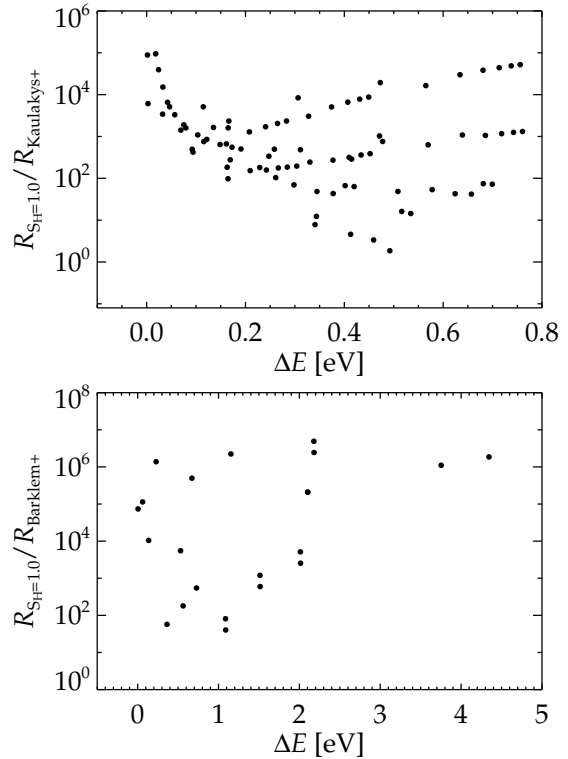}
          \caption{Comparison between rate coefficients at $T=6000$\,K
            for collisional excitation of Na\,I by neutral hydrogen
            atoms. Only optically allowed transitions are shown and
            x-axis represents the energy of the
            transition. \textit{Top:} The ratio between the unscaled
            Drawin formula and the free electron model of
            \citet{Kaulakys91} for transitions between highly excited
            states. \textit{Bottom:} The ratio between the unscaled
            Drawin formula and the quantum mechanical calculations by
            \citet{Barklem10} for transitions between levels up to
            5s.}
        \label{fig:hcoll}
\end{figure}

\citet{Belyaev10} have performed quantum-scattering calculations of
cross-sections for inelastic collisions between $\rm H+Na$ with impact
energies ranging from threshold up to 10\,eV. Using
these cross-sections, \citet{Barklem10} present for the first time quantum
mechanical rate coefficients 
for a large number of bound-bound transitions in Na\,I, caused by
collisions with neutral hydrogen, as well as charge exchange reactions
between sodium and hydrogen ($\rm Na^*+H \getsto Na^++H^-$). We use
the rate coefficients given in \citet{Barklem10} involving all levels and all
possible transitions below the ionic limit, i.e. up to 5p (for levels
above the ionic limit, charge exchange is not possibile).
The bound-bound rates for allowed transitions are smaller
than what is found with the commonly used classical recipe by Drawin
by one to six orders of magnitude, depending on the transition (see
Fig.\,\ref{fig:hcoll}). The Drawin recipe is not applicable to
forbidden bound-bound transitions, ionisation or charge exchange. 

We estimate the bound-bound collisional rates for transitions between
highly excited levels with the free electron model described by
\citet{Kaulakys91}, which is applicable to Rydberg atoms (Eq. 18 in 
\citeauthor{Kaulakys91}, using non-hydrogenic wave-functions in momentum
space calculated using the methods of
\citet{HoangBinh97}). Figure \ref{fig:hcoll} compares the rate
coefficients found by \citet{Barklem10} with those given by the Drawin
recipe for transitions between low excited states, and also compares
the rates by Kaulakys and Drawin for transitions between highly
excited states. It is easily appreciated from these figures that for
Na no single scaling factor can be applied to the Drawin recipe to reproduce
either the quantum mechanical rates or the Kaulakys rates.  
   
\citet{Barklem10} also provide estimates of the rate uncertainty, so
called fluctuation factors. In Sect.\,\ref{sec:influence} we discuss
the effect on
the non-LTE calculations when multiplying all quantum mechanical rates
by the maximum fluctuation factors or by the suggested minimum value
(0.5).

\subsection{Line broadening}
\label{sec:broad}
For line broadening caused by collisions with neutral hydrogen, we use
the ABO theory by \citet{Anstee95} and \citet{Barklem97} whenever applicable, and
otherwise adopt the $C_6$ constant for van der Waals-broadening by
\citet{Unsold55}. The latter are enhanced by a factor of 2
($\Delta\log{C_6}\approx0.8$), which is typically required to match
the observed line profiles in the Sun
\citep[e.g.][]{Mashonkina00}. Stark broadening constants ($C_4$) are
estimated from the tables of
\citet{Dimitrijevic85,Dimitrijevic90}. The line data are summarised in
Table \ref{tab:lines} for the 11 principally considered transitions. 

\section{Discussion}
\label{sec:discussion}

\begin{figure}
        \centering
                \includegraphics[angle=90,width=6.5cm,viewport=0cm 12cm 19cm 23cm]{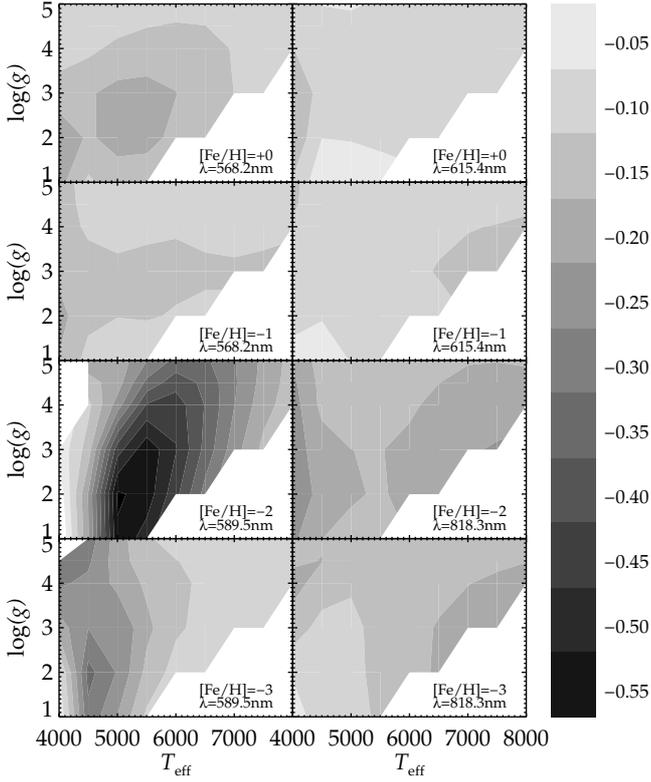}
          \caption{Contour diagrams illustrating how the abundance
            corrections vary with effective temperature and surface
            gravity for four commonly analysed lines of
            Na\,I. Only the results for $\xi=2$\,km/s and
            $\rm[Na/Fe]=0$ are shown.}
        \label{fig:cont}
\end{figure}

\begin{figure}
        \centering
                \includegraphics[angle=90,width=6.5cm,viewport=0cm 13cm
                19cm 24cm]{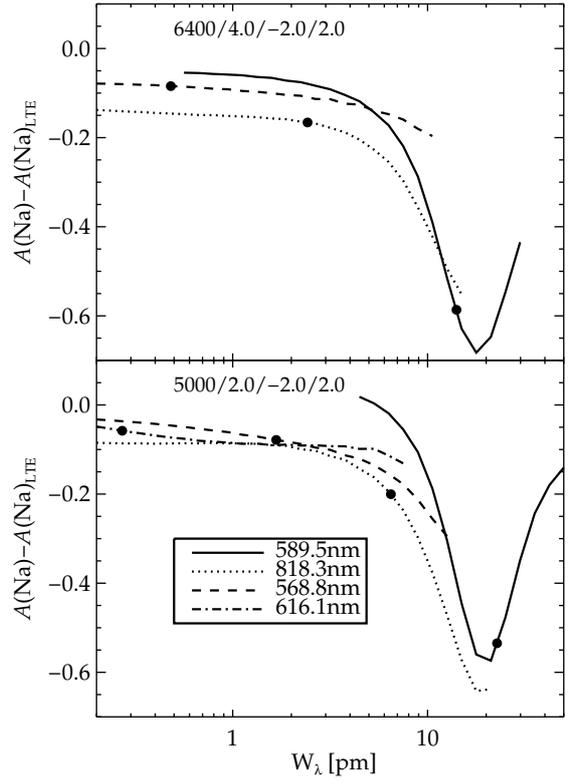}
          \caption{Non-LTE abundance corrections as functions of
            equivalent width for selected Na\,I lines for a metal-poor
            turn-off star (\textit{top:} $T_{\rm eff}=6400$\,K,
            $\log{g}=4.0$, $\rm[Fe/H]=-2$, and $\xi_{t}=2.0$\,km/s)
            and metal-poor giant (\textit{bottom:} $T_{\rm eff}=5000$\,K,
            $\log{g}=2.0$, $\rm[Fe/H]=-2$, and
            $\xi_{t}=2.0$\,km/s). The black bullets indicate the
            predicted line strength and abundance corrections at
            $\rm[Na/Fe]=0$}
        \label{fig:corr}
\end{figure}

\subsection{Departures from LTE}
\label{sec:deps}
There is a great regularity in the statistical equilibrium of Na over
the whole stellar grid covered by our analysis. As we will describe
below, the non-LTE abundance corrections are dependent on the line
strength and much of the model dependence can therefore be easily
understood through the variation of the excitation and ionisation
equilibrium with temperature and density. Figure \ref{fig:cont} shows
contour plots of how the inferred corrections vary with effective
temperature and surface gravity at $\rm[Na/Fe]=0$. 

To describe the non-LTE line formation of Na lines, we define the
departure coefficient $\beta_x=N_x/N^*_x$, which represents the population
of a certain level $x$ in non-LTE, divided by the corresponding
population in LTE. Given a specific background continuum opacity, the
strength and shape of a spectral line are determined by the line
opacity, which is proportional to the population of the lower level of the
transition, and the source function $S_\nu$, which is proportional to the
population ratio between the upper and lower level\footnote{Neglecting
  stimulated emission, the line source function $S_l$ is directly
  given by the expression $S_l/B_\nu=\beta_u/\beta_l$}. Lifting the
assumption of LTE, both line formation depth and source function
may change and thus alter the strength and shape of the spectral
line. 

The strong Na\,I D doublet lines at 588.9/589.5\,nm are commonly used
in abundance studies of metal-poor stars. Indeed, in certain cases
(warm, metal-poor dwarfs) they are the only lines that are
sufficiently strong. The doublet originates from the resonance line
transition between the ground state (3s) and the two fine structure
components of the first excited state ($\rm 3p_{1/2}$ and $\rm
3p_{3/2}$). We found empirically that as a general rule 
the line source function
perfectly resembles that of a pure scattering line in a two-level atom, i.e. to
a very good approximation $S_l=\bar{J}_\phi$ at all depths, where
$\bar{J}_\phi$ is the profile-averaged mean intensity. This merely
reflects that for these lines, the photon absorption and
emission rates strongly dominate the collisional rates and all
interactions with other levels, including the continuum. Although in
principle for a two-level atom,
$S_l=(1-\epsilon_\nu)\bar{J}_\phi+\epsilon_\nu B_\nu$, the probability
for true absorption, $\epsilon_\nu$, is very close to zero at shallow
depths where $\bar{J}_\phi$ departs from $B_\nu$, so the approximation
of a pure scattering line still holds at all depths. 

The ratio between the population of the upper and lower level is
therefore always set by the mean intensity, and, for a specific line
strength, this is correctly established with a simple two-level
atom. However, the actual population of the ground level, which
governs the line opacity and typical formation depth of the line, is
underestimated when more highly excited levels of Na\,I are
neglected. As described by \citet{Bruls92}, a number of
high-excitation levels and a ladder of high-probability transitions
connecting these with lower excitation levels must be established to
obtain the correct populations. This is needed because the first and
second excited states have photoionisation thresholds in the
ultra-violet, where the radiation field exceeds the Planck function
($J_\nu>B_\nu$), and pushes the ionisation balance to overionisation of
Na\,I. The situation is reversed for more highly excited levels since
their photoionisation thresholds lie in the near-infrared regime,
where the radiation field is instead subthermal. This
photoionisation/recombination picture has been described previously by
\citet{Takeda03} and \citet{Mashonkina00}.

We can now qualitatively understand the non-LTE formation, starting
with the Na\,I resonance lines. When either line is weak, it will
obviously have little influence on its own radiation field. Therefore,
$J_\nu-B_\nu>0$ at shallow optical depths, as is the case for
neighbouring continuum
regions. This is governed by the temperature gradient, which
determines how rapidly $B_\nu$ decreases with optical depth. At these
wavelengths, the gradient is steep enough to produce a $J_\nu-B_\nu$
excess and, consequently, a line source function that is stronger than
in LTE ($\beta_u/\beta_l>1$). This would tend to weaken the resonance lines
compared with LTE, but the effect is small because the lines are formed deep in the
atmosphere, where the radiation field is close to thermal. 
On the other hand, the photoionisation balance of Na\,I is
always shifted to overrecombination in the line-forming regions. The
ground state of Na\,I is thus overpopulated compared with LTE
($\beta_l>1$), and the combined effect is generally a moderate line
strengthening. The resulting abundance corrections for the resonance
lines are always minor, $|\Delta A|<0.1$\,dex at line strengths below
5\,pm. However, these lines are only that weak in extremely metal-poor
stars. 

As the line strength increases towards saturation in any given model,
$J_\nu$ drops, and the line source function becomes weaker than in
LTE. This is naturally accompanied by a decrease in the excitation
rate. Also the ionisation rates drop with increasing abundance as the
photoionising radiation field weakens. This enhances photorecombination from
the Na\,II reservoir, and all levels of Na\,I are increasingly
overpopulated. The combination of higher line opacity and a weaker
source function produce significant line strengthening in non-LTE,
with negative abundance corrections as a result. This behaviour is
illustrated in Fig.\,\ref{fig:corr} for a metal-poor dwarf and giant. 
We note
that even if the statistical equilibrium is not necessarily pushed
further from LTE, the abundance corrections become larger and larger
as the line saturates, simply because the abundance sensitivity to
equivalent width is small in this regime. 

With further strengthening, the line enters onto the damping part of
the curve-of-growth with the development of broad wings. Because photons
from a wider frequency range are then able to excite the atoms, the
excitation rate actually increases again, lessening the
overpopulation in deep layers. The abundance sensitivity to line
strength also starts to increase again, consequently the abundance corrections,
as we define them here, reach a minimum value when the line is fully
saturated (see Fig.\,\ref{fig:corr}) and then become less negative
with higher abundance, although these strong lines are hardly suitable
for accurate abundance analysis through equivalent width
measurements. Line profile analysis especially of the wings is more 
appropriate, but that is not addressed here.

Even if the two-level approximation holds true only for the source
function of the resonance
lines, the subordinate transitions (with 3p as lower level) show a
very similar behaviour. At low line strength each line has a 'plateau'
of close-to-constant, small abundance corrections, which becomes
increasingly negative as the line saturates around 20\,pm. This
general behaviour with line strength has been discovered previously
e.g. by \citet{Takeda03}. The strongest subordinate transitions,
3p--3d at 818.3/819.4\,nm, follow an almost identical behaviour as the
resonance lines, but are offset to more negative corrections. The
latter is because of the shallower temperature gradient with continuum optical depth
in the near-infrared wavelength, which produces a $J_\nu-B_\nu$ deficiency
in the line, and a source function that is weaker than in LTE also
when the lines are very weak. 

Naturally, the abundance corrections at a given line strength are
still somewhat model-dependent. For saturated lines, the corrections
are more negative for hotter models and models with lower surface
gravity, whereas the metallicity dependence seems almost
negligible. Fully unsaturated lines (below 5\,pm) almost always have
corrections in the range $-0.1...-0.2$\,dex.
 
We note the very close resemblance in the non-LTE line
formation of sodium and the lighter alkali atom lithium, whose
departures from LTE have been described e.g by 
\citet{Carlsson94} and \citet{Lind09a}. There are many striking
similarities between the
two elements, especially in the shape of the abundance correction
curves. Differences mainly arise from the higher degree of
overionisation of Li, which in turn is a direct result of the larger
photoionisation cross-section from the first excited state of Li\,I (2p),
compared to Na\,I (3p). The abundance corrections at low line
strengths thus tend to be somewhat higher, even positive, for Li
(overionisation causes underpopulation of Li\,I, thus weakening the
spectral lines). 

\subsection{The influence of collisions}
\label{sec:influence}
As described in Sects.\,\ref{sec:ecoll} and \ref{sec:hcoll}, the
statistical equilibrium of Na\,I
was calculated by accounting for collisional excitation and ionisation
by electrons and hydrogen atoms. We now discuss the impact on the
derived abundance corrections when varying the strength of collisional
rates.

When multiplying/dividing all rates for collisional excitation and
ionisation by electrons by a factor of two, the solar equivalent
widths of Na\,I lines change systematically by typically 1\%. This
propagates to less than $\sim0.01$\,dex in terms of non-LTE abundance
corrections, and is therefore negligible. A somewhat larger impact
is seen for hotter, higher surface gravity models, where electrons are
more abundant. Still, the non-LTE equivalent widths calculated for a
$T_{\rm eff}=8000$\,K, log$(g)=5.0$, solar metallicity model are
affected by only 2--4\%, corresponding to approximately 0.02\,dex for
relevant lines. Giants and cooler dwarfs are less sensitive, because
they have 
lower densities of the colliding free electrons. The Na non-LTE
calculations thus seem robust with respect to input atomic data for
collisions with electrons.     

\begin{figure}
        \centering
                \includegraphics[angle=90,width=8.5cm]{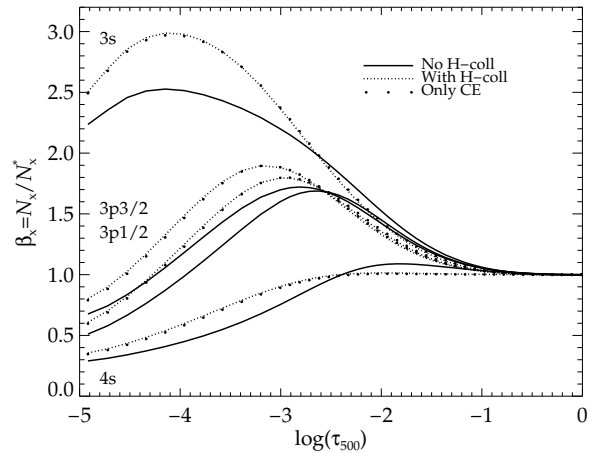}
          \caption{Departure coefficients for a solar model, for the
            Na\,I levels indicated with labels in the
            plot. \textit{Solid:} neglecting collisions with neutral
            hydrogen. \textit{Dotted:} including charge exchange
            reactions and bound-bound collisional excitation by
            neutral hydrogen. \textit{Bullets:} including only charge
            exchange reactions in collisions with hydrogen.}
        \label{fig:deps}
\end{figure}

The quantum mechanical calculations by \citet{Barklem10} result in
rate coefficients for collisional excitation by hydrogen atoms that
are lower than the commonly used classical Drawin recipe by one to
six orders of magnitude. The new rates thus have very little influence
on the statistical equilibrium of Na\,I, in the stellar parameter
range that we consider here. However, again analogously to Li, charge
exchange reactions turn out to be more
influential. Figure \ref{fig:deps} illustrates how the departure
coefficients of low excited levels change for a solar model, when
including and neglecting collisions with neutral hydrogen. 

It is conceptually correct to say that higher collisional rates have a
thermalising effect, i.e. they tend to drive the level population towards
LTE. However, as seen for the Sun in Fig.\,\ref{fig:deps}, this is not
generally true for all levels. The populations of the ground state and
first excited states are instead pushed farther away from LTE in optically
thin atmospheric layers, when charge exchange is included. This
seemingly contradictory behaviour can be understood by realising that
the collisional cross-sections of 3s and 3p are very small, and the
changes in their level population are rather a secondary effect,
stemming from the thermalisation of higher excited levels. Especially
4s, whose departure coefficients are also displayed in
Fig.\,\ref{fig:deps}, has a large cross-section for charge exchange
and the level becomes almost thermalised with the inclusion of this
process. In deeper layers, where $\beta_{\rm 4s}>1$, the
overpopulation of the level decreases, which also propagates to 
lower excited levels. In
shallow layers, where $\beta_{\rm 4s}<1$, the added neutralisation
channel through charge exchange reactions lessens the underpopulation
of 4s, but further increases the
overpopulation of lower levels. In practice, higher excited levels
than 4s influence the outcome as well, but to a lesser extent.

Even if the statistical equilibrium is indeed influenced by hydrogen
collisions, this need not necessarily be the case for the 
the emergent spectral lines. This is because the source functions
remain unchanged, and as discussed above, the effect on the level
populations of 3s and 3p is opposite in different regimes of the
atmosphere, so that the net effect on the line strength is small. The
situation is different from that of Li, for which charge exchange
reactions always lead to more neutralisation throughout the atmosphere,
with significant line strengthening as a result
\citep{Barklem03b,Lind09a}. 

Charge exchange is most important in cool, high surface gravity
models, for which abundance corrections are affected by up to
$\sim$0.2\,dex, whereas the effect in solar-type stars is only
$\sim$0.01\,dex. This is a direct reflection of the ionisation
equilibrium of hydrogen, because H$^-$ is abundant in cool, dense
atmospheres. We therefore tested how the uncertainty in
collisional rates affects the abundance corrections for the extreme
case of a $T_{\rm eff}=4000$\,K, $\log{g}=5.0$ model. When multiplying
and dividing all rates with their maximum fluctuation factors (see
Sect.\,\ref{sec:hcoll}), the abundance corrections vary with typically
0.04\,dex
for such a model. For the Sun, the corresponding variation is
$<0.01\,$dex, i.e. essentially irrelevant for practical purposes.

\begin{figure}
        \centering \includegraphics[angle=90,width=9.0cm,viewport=1cm
          14cm 20cm 25cm]{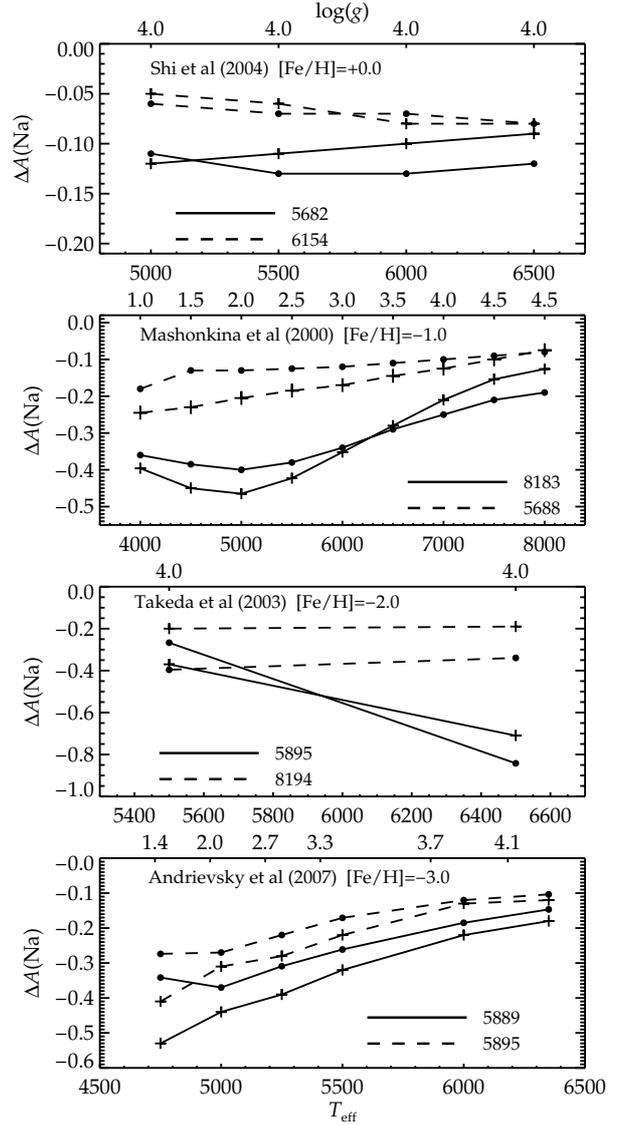}
          \caption{Comparison between the abundance corrections
            determined in this study (lines connected with bullets)
            and earlier studies, as indicated in each panel (lines
            connected with plus signs). Effective temperature is used
            as reference axis, with surface gravity values for each
            model indicated along the top axis of each panel. All
            calculations assumed $\rm[Na/Fe]=0.0$.}
        \label{fig:other}
\end{figure}

\subsection{Consequences for stellar abundance analysis}
\label{sec:conseq}
Obviously, it is always preferable to base the abundance analysis on
unsaturated spectral lines, not least because non-LTE abundance
corrections for fully saturated lines may reach $-1$\,dex in extreme
cases. This means that equivalent width measurements of 
the resonance lines should be used only as a
last resort for metal-poor stars, when the
818.3/819.4\,nm doublet lines are not detectable or too severely
blended with telluric lines. At metallicities in the range
$\rm[Fe/H]=-2.0...-1.0$, the 568.2/568.8\,nm doublet lines are usually
to be preferred, whereas the 615.4/616.0 are good indicators for solar
metallicity stars. Given the estimated uncertainties in collisional
data, the non-LTE modelling gives abundances for weak lines that
normally can be trusted to high precision.

Figure \ref{fig:other} compares our inferred corrections with those of
\citet{Andrievsky07}, \citet{Shi04}, \citet{Takeda03} and
\citet{Mashonkina00} for selected models within the grid. Overall, the
agreement between the studies is never worse than 0.2\,dex and they 
usually agree to within 0.1\,dex. Considering that all studies
have (to various degree) differences in input atomic data, atmospheric
models, and the numerical methods, this is not surprising. Still, all
statistical equilibrium calculations in 1D are conceptually similar
and a better consensus between independent studies is clearly desired
to ensure that accurate Na abundances can be derived. In particular, 
we note the sometimes highly significant discrepancies with 
the non-LTE calculations published by \citet{Gratton99}. Their
abundance corrections show an increasing trend toward
positive values (up to $+0.5$\,dex, depending on the line) for very
low surface gravity stars, which is not seen in our calculations. 

\begin{figure}
        \centering \includegraphics[angle=90,width=9.0cm,viewport=0cm
          15cm 19.5cm 25.5cm]{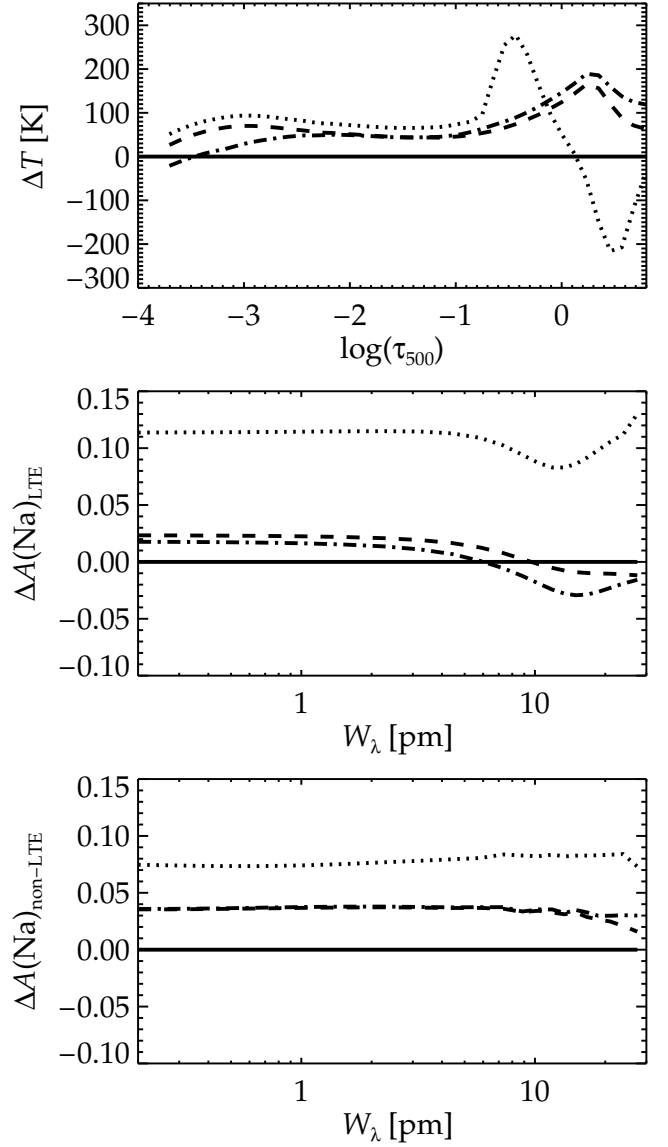}
          \caption{\textit{Top:} Temperature difference between \textsc{ATLAS9}
            models with respect to \textsc{MARCS} models \citep{Gustafsson08}, for
            a metal-poor giant ($T_{\rm eff}=5000$\,K, $\log{g}=2.0$,
            $\rm [Fe/H]=-2.0$, $\xi_t=2.0$\,km/s). The dotted line
            represents an \textsc{ATLAS9} model with scaled solar composition
            and convective overshoot from the Kurucz 1993-grid. The
            dashed line has the same composition, but without
            overshoot. The dashed-dotted line represents a 0.4\,dex
            alpha-enhanced \textsc{ATLAS9} model without overshoot from the
            2004-grid by Castelli \& Kurucz. \textit{Middle:}
            Difference between LTE abundance as function of line
            strength for the same models with respect to \textsc{MARCS}.
            The abundances are based on the 819.4\,nm Na\,I line 
            and vary from left to right between
            $\rm[Na/Fe]=-2.0...+2.0$. \textit{Bottom:} Same as the middle
            panel, but for non-LTE abundances.}
        \label{fig:markur}
\end{figure}

While our model atom is superior to previous studies in that it
contains more reliable atomic data for collisions, we do not expect
that this is the primary reason behind some of the larger offsets
compared with
previous studies. As discussed in Sect.\,\ref{sec:influence}, the
sensitivity to the
detailed collisional rates is fairly low for the specific case of Na
in the stellar parameter range covered and we suspect that the differences
in the model atmospheric structure have a larger impact. \citet{Takeda03},
\citet{Mashonkina00} and \citet{Andrievsky07} all use Kurucz \textsc{ATLAS9}
models, while \citet{Shi04} use the grid by \citet{Fuhrmann97}. The
atmospheric temperature gradient is inevitably different for models of
identical stellar parameters, for example because of differences in
the treatment of convection, equation of state, and opacity. This is
illustrated in the top panel of Fig.\,\ref{fig:markur}, which shows the
difference in
temperature stratification for different versions of \textsc{ATLAS9} models
with respect to a \textsc{MARCS} model for a metal-poor giant\footnote{The \textsc{ATLAS9}
  models were retrieved from \url{http://kurucz.harvard.edu/grids/},
  with convective overshoot (gridm20), without overshoot
  (gridm20NOVER), and using more up-to-date opacities and abundances
  (gridm20a2ODFNEW).}. As can be seen in this figure, the
stratification is critically dependent on the inclusion of convective
overshooting in the 1993 grid by Kurucz. The most up-to-date versions
of \textsc{ATLAS9} \citep{Castelli04} and \textsc{MARCS} \citep{Gustafsson08} both
neglect overshoot, and the models shown have similar, but not
identical parameters for the mixing length, but nevertheless show
differences on the 100\,K level.

The middle and lower panels of Fig.\,\ref{fig:markur} illustrate the effect on
derived LTE and non-LTE Na abundances for the 819.4\,nm line, when
using different atmospheric models. In both cases the line strengths
appear, as expected, to be connected to the
temperatures around $\log(\tau_{500})=0$. The higher temperatures of
the \textsc{ATLAS9} models, especially the one with overshoot, weakens the line
compared to \textsc{MARCS} for a given Na abundance (i.e. a given line strength
generally corresponds to a higher LTE abundance). However, in non-LTE
the difference is lessened because of a higher degree of
overpopulation of neutral Na in the temperature 'jump' in the overshooting
model. This in turn is caused by the flatter temperature gradient and
consequently the less efficient overionising ultra-violet flux in this region. 
This comparison implies that differences between 1D atmospheric models 
can produce differences of the order of $\sim0.1$\,dex in Na abundance
for such a metal-poor giant.

\section{Concluding remarks}
\label{sec:remarks}
Using our calculated abundance corrections and/or non-LTE
curves-of-growth for 11 important neutral Na\,I lines, Na abundances
that are superiour to those inferred with the LTE assumption can
easily be obtained for late-type dwarf and giant stars. The results
show a low sensitivity to uncertainties in input atomic data for
collisional cross-sections, but are sensitive to the detailed
structure of the atmosphere. To minimize the influence from possible
systematic errors in the model, unsaturated lines are definitely to be
preferred as abundance indicators. As always, a good test of the
soundness of the modelling procedure is to compare Na abundances from
lines of different strengths.

Our results satisfactorily agree with earlier non-LTE studies
of Na\,I that treated the strength of hydrogen collisions as a free
parameter, even though
the new quantum mechanical rates for hydrogen impact excitation are
considerably lower than the traditational estimates. This low
sensitivity of non-LTE abundances to hydrogen collisions stems from a
cancellation between effects in different line-forming regions.

In a forthcoming paper we will explore the impact of the model
atmosphere on the abundance determination more closely, and extend our
work to include 3D, hydrodynamical model atmospheres, superseding the
crude mixing length recipes used for convection in static 1D
models. The model atom presented here will then be applied to specific
cases of solar-metallicity and metal-poor dwarfs and giants, and we
will discuss the Galactic chemical evolution of Na.

\acknowledgements{
We gratefully acknowledge the support of the Royal Swedish Academy of
Sciences, G{\"o}ran Gustafssons Stiftelse and the Swedish Research
Council.  P.~S.~B is a Royal Swedish Academy of Sciences Research Fellow
supported by a grant from the Knut and Alice Wallenberg
Foundation. A.~K.~B. acknowledges the support from the Russian
Foundation for Basic Research (Grant No. 10-03-00807-a). Special
thanks to N. Feautrier for assistance with the draft and
for providing us with collisional data before publication.}


\begin{thebibliography}{47}
\expandafter\ifx\csname natexlab\endcsname\relax\def\natexlab#1{#1}\fi

\bibitem[{Allen(2000)}]{Allen00}
Allen, C.~W. 2000, {Allen's Astrophysical Quantities}, 4th edn. (Springer,
  Berlin)

\bibitem[{{Andrievsky} {et~al.}(2007){Andrievsky}, {Spite}, {Korotin}, {Spite},
  {Bonifacio}, {Cayrel}, {Hill}, \& {Fran{\c c}ois}}]{Andrievsky07}
{Andrievsky}, S.~M., {Spite}, M., {Korotin}, S.~A., {et~al.} 2007, \aap, 464,
  1081

\bibitem[{{Anstee} \& {O'Mara}(1995)}]{Anstee95}
{Anstee}, S.~D. \& {O'Mara}, B.~J. 1995, \mnras, 276, 859

\bibitem[{{Asplund}(2005)}]{Asplund05}
{Asplund}, M. 2005, \araa, 43, 481

\bibitem[{{Barklem} {et~al.}(2003){Barklem}, {Belyaev}, \&
  {Asplund}}]{Barklem03b}
{Barklem}, P.~S., {Belyaev}, A.~K., \& {Asplund}, M. 2003, \aap, 409, L1

\bibitem[{{Barklem} {et~al.}(2010){Barklem}, {Belyaev}, {Dickinson}, \&
  {Gad{\'e}a}}]{Barklem10}
{Barklem}, P.~S., {Belyaev}, A.~K., {Dickinson}, A.~S., \& {Gad{\'e}a}, F.~X.
  2010, \aap, 519, A20+

\bibitem[{{Barklem} \& {O'Mara}(1997)}]{Barklem97}
{Barklem}, P.~S. \& {O'Mara}, B.~J. 1997, \mnras, 290, 102

\bibitem[{{Belyaev} {et~al.}(2010){Belyaev}, {Barklem}, {Dickinson}, \&
  {Gad{\'e}a}}]{Belyaev10}
{Belyaev}, A.~K., {Barklem}, P.~S., {Dickinson}, A.~S., \& {Gad{\'e}a}, F.~X.
  2010, \pra, 81, 032706

\bibitem[{{Bensby} {et~al.}(2003){Bensby}, {Feltzing}, \&
  {Lundstr{\"o}m}}]{Bensby03}
{Bensby}, T., {Feltzing}, S., \& {Lundstr{\"o}m}, I. 2003, \aap, 410, 527

\bibitem[{{Bruls} {et~al.}(1992){Bruls}, {Rutten}, \& {Shchukina}}]{Bruls92}
{Bruls}, J.~H.~M.~J., {Rutten}, R.~J., \& {Shchukina}, N.~G. 1992, \aap, 265,
  237

\bibitem[{{Carlsson}(1986)}]{Carlsson86}
{Carlsson}, M. 1986, Uppsala Astronomical Observatory Reports, 33

\bibitem[{{Carlsson}(1992)}]{Carlsson92}
{Carlsson}, M. 1992, in Astronomical Society of the Pacific Conference Series,
  Vol.~26, Cool Stars, Stellar Systems, and the Sun, ed. M.~S. {Giampapa} \&
  J.~A. {Bookbinder}, 499--+

\bibitem[{{Carlsson} {et~al.}(1994){Carlsson}, {Rutten}, {Bruls}, \&
  {Shchukina}}]{Carlsson94}
{Carlsson}, M., {Rutten}, R.~J., {Bruls}, J.~H.~M.~J., \& {Shchukina}, N.~G.
  1994, \aap, 288, 860

\bibitem[{{Carretta} {et~al.}(2009){Carretta}, {Bragaglia}, {Gratton}, \&
  {Lucatello}}]{Carretta09b}
{Carretta}, E., {Bragaglia}, A., {Gratton}, R., \& {Lucatello}, S. 2009, \aap,
  505, 139

\bibitem[{{Castelli} \& {Kurucz}(2004)}]{Castelli04}
{Castelli}, F. \& {Kurucz}, R.~L. 2004, ArXiv Astrophysics e-prints

\bibitem[{{Cayrel} {et~al.}(2004){Cayrel}, {Depagne}, {Spite}, {Hill}, {Spite},
  {Fran{\c c}ois}, {Plez}, {Beers}, {Primas}, {Andersen}, {Barbuy},
  {Bonifacio}, {Molaro}, \& {Nordstr{\"o}m}}]{Cayrel04}
{Cayrel}, R., {Depagne}, E., {Spite}, M., {et~al.} 2004, \aap, 416, 1117

\bibitem[{{Cunto} \& {Mendoza}(1992)}]{Cunto92}
{Cunto}, W. \& {Mendoza}, C. 1992, \rmxaa, 23, 107

\bibitem[{{Denisenkov} \& {Denisenkova}(1990)}]{Denisenkov90}
{Denisenkov}, P.~A. \& {Denisenkova}, S.~N. 1990, Soviet Astronomy Letters, 16,
  275

\bibitem[{{Dimitrijevi{\'c}} \& {Sahal-Br{\'e}chot}(1985)}]{Dimitrijevic85}
{Dimitrijevi{\'c}}, M.~S. \& {Sahal-Br{\'e}chot}, S. 1985, Journal of
  Quantitative Spectroscopy and Radiative Transfer, 34, 149

\bibitem[{{Dimitrijevi{\'c}} \& {Sahal-Br{\'e}chot}(1990)}]{Dimitrijevic90}
{Dimitrijevi{\'c}}, M.~S. \& {Sahal-Br{\'e}chot}, S. 1990, Journal of
  Quantitative Spectroscopy and Radiative Transfer, 44, 421

\bibitem[{{Drawin}(1968)}]{Drawin68}
{Drawin}, H.-W. 1968, Zeitschrift fur Physik, 211, 404

\bibitem[{{Edvardsson} {et~al.}(1993){Edvardsson}, {Andersen}, {Gustafsson},
  {Lambert}, {Nissen}, \& {Tomkin}}]{Edvardsson93}
{Edvardsson}, B., {Andersen}, J., {Gustafsson}, B., {et~al.} 1993, \aap, 275,
  101

\bibitem[{{Fuhrmann} {et~al.}(1997){Fuhrmann}, {Pfeiffer}, {Frank}, {Reetz}, \&
  {Gehren}}]{Fuhrmann97}
{Fuhrmann}, K., {Pfeiffer}, M., {Frank}, C., {Reetz}, J., \& {Gehren}, T. 1997,
  \aap, 323, 909

\bibitem[{{Gao} {et~al.}(2010){Gao}, {Han}, {Voky}, {Feautrier}, \&
  {Li}}]{Gao10}
{Gao}, X., {Han}, X., {Voky}, L., {Feautrier}, N., \& {Li}, J. 2010, \pra, 81,
  022703

\bibitem[{{Gehren} {et~al.}(2006){Gehren}, {Shi}, {Zhang}, {Zhao}, \&
  {Korn}}]{Gehren06}
{Gehren}, T., {Shi}, J.~R., {Zhang}, H.~W., {Zhao}, G., \& {Korn}, A.~J. 2006,
  \aap, 451, 1065

\bibitem[{{Gratton} {et~al.}(2004){Gratton}, {Sneden}, \&
  {Carretta}}]{Gratton04}
{Gratton}, R., {Sneden}, C., \& {Carretta}, E. 2004, \araa, 42, 385

\bibitem[{{Gratton} {et~al.}(1999){Gratton}, {Carretta}, {Eriksson}, \&
  {Gustafsson}}]{Gratton99}
{Gratton}, R.~G., {Carretta}, E., {Eriksson}, K., \& {Gustafsson}, B. 1999,
  \aap, 350, 955

\bibitem[{{Grevesse} {et~al.}(2007){Grevesse}, {Asplund}, \&
  {Sauval}}]{Grevesse07}
{Grevesse}, N., {Asplund}, M., \& {Sauval}, A.~J. 2007, \ssr, 130, 105

\bibitem[{{Gustafsson} {et~al.}(2008){Gustafsson}, {Edvardsson}, {Eriksson},
  {J{\o}rgensen}, {Nordlund}, \& {Plez}}]{Gustafsson08}
{Gustafsson}, B., {Edvardsson}, B., {Eriksson}, K., {et~al.} 2008, \aap, 486,
  951

\bibitem[{{Hoang Binh} \& {Van Regemorter}(1997)}]{HoangBinh97}
{Hoang Binh}, D. \& {Van Regemorter}, H. 1997, Journal of Physics B Atomic
  Molecular Physics, 30, 2403

\bibitem[{{Igenbergs} {et~al.}(2008){Igenbergs}, {Schweinzer}, {Bray}, {Bridi},
  \& {Aumayr}}]{Igenbergs08}
{Igenbergs}, K., {Schweinzer}, J., {Bray}, I., {Bridi}, D., \& {Aumayr}, F.
  2008, Atomic Data and Nuclear Data Tables, 94, 981

\bibitem[{{Kaulakys}(1991)}]{Kaulakys91}
{Kaulakys}, B. 1991, Journal of Physics B Atomic Molecular Physics, 24, L127

\bibitem[{{Lind} {et~al.}(2009){Lind}, {Asplund}, \& {Barklem}}]{Lind09a}
{Lind}, K., {Asplund}, M., \& {Barklem}, P.~S. 2009, \aap, 503, 541

\bibitem[{{Lind} {et~al.}(submitted){Lind}, {Charbonnel}, {Decressin},
  {Primas}, {Grundahl}, \& {Asplund}}]{Lind10a}
{Lind}, K., {Charbonnel}, C., {Decressin}, T., {et~al.} submitted, \aap

\bibitem[{{Mashonkina} {et~al.}(2000){Mashonkina}, {Shimanski{\u i}}, \&
  {Sakhibullin}}]{Mashonkina00}
{Mashonkina}, L.~I., {Shimanski{\u i}}, V.~V., \& {Sakhibullin}, N.~A. 2000,
  Astronomy Reports, 44, 790

\bibitem[{{Nissen} \& {Schuster}(2010)}]{Nissen10}
{Nissen}, P.~E. \& {Schuster}, W.~J. 2010, \aap, 511, L10+

\bibitem[{{Park}(1971)}]{Park71}
{Park}, C. 1971, Journal of Quantitative Spectroscopy and Radiative Transfer,
  11, 7

\bibitem[{{Phelps} \& {Lin}(1981)}]{Phelps81}
{Phelps}, J.~O. \& {Lin}, C.~C. 1981, \pra, 24, 1299

\bibitem[{{Reddy} {et~al.}(2003){Reddy}, {Tomkin}, {Lambert}, \& {Allende
  Prieto}}]{Reddy03}
{Reddy}, B.~E., {Tomkin}, J., {Lambert}, D.~L., \& {Allende Prieto}, C. 2003,
  \mnras, 340, 304

\bibitem[{{Sansonetti}(2008)}]{Sansonetti08}
{Sansonetti}, J.~E. 2008, Journal of Physical and Chemical Reference Data, 37,
  1659

\bibitem[{{Seaton}(1962)}]{Seaton62a}
{Seaton}, M.~J. 1962, Proceedings of the Physical Society, 79, 1105

\bibitem[{{Shi} {et~al.}(2004){Shi}, {Gehren}, \& {Zhao}}]{Shi04}
{Shi}, J.~R., {Gehren}, T., \& {Zhao}, G. 2004, \aap, 423, 683

\bibitem[{{Sydoryk} {et~al.}(2008){Sydoryk}, {Bezuglov}, {Beterov}, {Miculis},
  {Saks}, {Janovs}, {Spels}, \& {Ekers}}]{Sydoryk08}
{Sydoryk}, I., {Bezuglov}, N.~N., {Beterov}, I.~I., {et~al.} 2008, \pra, 77,
  042511

\bibitem[{{Takeda} {et~al.}(2003){Takeda}, {Zhao}, {Takada-Hidai}, {Chen},
  {Saito}, \& {Zhang}}]{Takeda03}
{Takeda}, Y., {Zhao}, G., {Takada-Hidai}, M., {et~al.} 2003, Chinese Journal of
  Astronomy and Astrophysics, 3, 316

\bibitem[{{Uns{\"o}ld}(1955)}]{Unsold55}
{Uns{\"o}ld}, A. 1955, {Physik der Sternatmospharen, MIT besonderer
  Berucksichtigung der Sonne.} (Berlin, Springer, 1955.~2.~Aufl.)

\bibitem[{{van Regemorter}(1962)}]{VanRegemorter62}
{van Regemorter}, H. 1962, \apj, 136, 906

\bibitem[{{Woosley} \& {Weaver}(1995)}]{Woosley95}
{Woosley}, S.~E. \& {Weaver}, T.~A. 1995, \apjs, 101, 181

\end{thebibliography}
\end{document}